\documentstyle[prl,aps,preprint]{revtex}

\begin{document}
\tighten

\title{LOCALITY OF THE STRANGE SEA IN THE NUCLEON\thanks
{This work is supported in part by funds provided by the U.S.
Department of Energy (D.O.E.) under cooperative agreement
\#DF-FC02-94ER40818.}}

\author{Xiangdong Ji and Jian Tang}

\address{Center for Theoretical Physics \\
Laboratory for Nuclear Science \\ and Department of Physics \\
Massachusetts Institute of Technology \\ Cambridge, Massachusetts
02139 \\ {~}}

\date{MIT-CTP-2456 \hskip 1in hep-ph/9507465 \hskip 1in July 1995}

\maketitle

\begin{abstract}
We introduce the concept of ``locality" for the strange sea
in the nucleon, which measures proximity of the strange
and anti-strange quarks in the momentum and coordinate spaces.
The CCFR data for the strange and anti-strange distributions
imply a ``local" strange sea in the momentum space, which
is unexpected in QCD and is at variance with
the simple meson-cloud model where the strangeness
is generated from the virtual transition of the nucleon to
a hyperon plus a kaon. We present a simple model to
interpret the CCFR data and to correlate
momentum and coordinate space locality,
yielding an upper bound of 0.005 fm$^2$ on
the strange radius. We also discuss significances of
locality for other
charge-conjugation-odd observables.

\end{abstract}

\pacs{xxxxxx}

\narrowtext

One of the fundamental properties of
quantum field theory is the existence of the Dirac
sea---infinite many fermions filling the negative
energy states predicted by the Dirac equation.
The excitation of the sea in the form of electron-positron
pairs is partly responsible for the famous
Lamb shift in the hydrogen spectra. Although novel,
the sea effects are essentially perturbative in
Quantum Electrodynamics because the fine
structure constant is small.
In Quantum Chromodynamics (QCD), however, the
quark sea plays a much more important role
in shaping the hadron structures due to the
strong coupling. For instance, the recent measurements of the
$g_1$ sum rule \cite{emc} indicate
that the quark sea is strongly polarized in
a polarized nucleon. Furthermore, the data from
deep-inelastic scattering show that the nucleon structure
functions are dominated by the sea quarks at
small Feynman $x$ \cite{sloan}.

Despite its importance, the detailed structure of the
quark sea in the
nucleon, is largely unknown. Because of this,
it is important to measure
as many as possible observables that
are directly linked to the sea degrees of freedom.
Recently there have been many discussions in the
literature about the ``strange content''
of the nucleon \cite{musolf}. These
discussions are mainly motivated
by the observation that strange quarks in the
nucleon, if there are any, must be generated from
the excitation of the Dirac sea. One hopes
that once enough information is collected
about the sea,  a consistent picture may
emerge to account for all
properties measured.

In this paper, we discuss
one particular aspect of the quark sea: the
difference between quark and antiquark
distributions in momentum and coordinate spaces.
For up and down flavors, there is no
physical way to separate the valence
and sea quark contributions
to any physical observables, other than
{\it defining} the sea-quark distribution to be
that of the antiquarks. On the other hand,
for the strange flavor, both quark and anti-quark
distributions can be measured separately.
Apart from the constraint that
the total number of strange quarks must be
equal to that of strange antiquarks,
the two distributions do not have to be the same.
The difference provides a unique
window to the structure of the sea.

Depending upon interactions between quarks,
there exist two extreme limits. In the first limit,
quarks and antiquarks in the sea have exactly
the same spatial or momentum wavefunctions. This can happen
if they are tightly bound in pairs, or if they move
independently but undergo similar interactions
with other spectators. [We defer a discussion about the
spin part of the wavefunctions to the latter part of the
paper.]
In the second limit, the quark and antiquark
wavefunctions are qualitatively different from
each other. This happens when
quarks and antiquarks in the sea experience
different interactions and move
independently inside the nucleon. To differentiate
the two limits, we introduce the concept of ``locality
of the sea'', which measures the
average local similarity of quark and antiquark wavefunctions
in coordinate or momentum spaces.

We are going to discuss two observables that have
information one locality of the sea.
If $s(\vec r)$ and
$\bar s (\vec r)$ are spatial wavefunctions for
the $s$ and $\bar s$ quarks in the nucleon,
the strangeness radius,
\begin{equation}
\langle r^2 \rangle_s = \int d{\vec r} r^2
(|s(\vec r)|^2-|\bar s (\vec r)|^2) \ ,
\end{equation}
is a measure of locality in the coordinate space.
On the other hand, if $s(x)$ and $\bar s (x)$
are the strange quark and antiquark distributions,
where $x$
is the momentum fraction of the nucleon carried
by quarks in the infinite momentum frame, the
positive-definite moment,
\begin{equation}
  L_s = \int_0^1 |s(x)-\bar s (x)| dx \ ,
\end{equation}
is a measure of locality in momentum space.
Clearly, a large $\langle r^2 \rangle_s$ or $L_s$ corresponds
to small locality and a small $\langle r^2 \rangle_s$ or $L_s$ to
large locality.

The remainder of the paper is organized as follows.
We first comment on general expectations
about the sea quark interactions
in QCD. We then discuss the meson-cloud model \cite{sull} for the
strange sea, which seems to reflect the above
expectations\cite{thomas}.
We compare the model predictions with the
strange quark and antiquark distributions
extracted recently in neutrino deep-inelastic scattering by the CCFR
collaboration\cite{ccfr}. The qualitative
difference motivated us to consider a new model
for the strange sea, in which the
strange and anti-strange quarks are perturbatively
generated from the gluon splitting $a$ $la$
Altarelli and Parisi \cite{ap}. The
interactions of $s$ and $\bar s$ with
other quarks present in the nucleon
are modelled by effective masses. After that, we
attempt to correlate locality in momentum
and coordinate spaces,
giving a prediction for the strange radius.
Finally, we comment on the effects of locality in
other observables.

In extracting sea distributions from hard processes,
one usually makes the assumption that the sea quark distribution
is equal to the sea antiquark distribution.
This, however, is unfounded in QCD. Although charge-conjugation
symmetry says that the quark distribution in a nucleon
is equal to the antiquark distribution in an antinucleon,
there is no symmetry in QCD lagrangian that allows to relate
quark and antiquark distributions of the nucleon in the sea. At the level
of Feynman diagrams, one can find a corresponding interaction of
the sea antiquarks for every interaction of the sea quarks
by changing the direction of quark propagation.
However, the strengths of the interactions
are different due to different color factors associated with
directions of the quark lines. An example is shown in Fig. 1,
where two diagrams are different only in the fermion-number
flow of the loop. It is easy to see that the color factors
from the two diagrams are quite different (signs are opposite).
In fact, in the large $N_c$ limit,
the quark interaction is suppressed relative to the
antiquark one by a factor $1/N_c^2$. When similar
diagrams with different topology  are taken into account, it
turns out that the quark and antiquark interactions have
the same order in $1/N_c$. However, there is no constraint
in QCD that they must have the same numerical coefficients.

If quarks and antiquarks in the sea indeed have significantly
different interactions, the meson cloud model is a good
representative of this. According to the model, the strange
sea is generated from the dissociation of the nucleon into
hyperons ($\Sigma$ and $\Lambda$) plus kaons. Thus, the strange
quark, mainly interacting with one up and one down quarks in
$\Lambda$, gives rise to the $s(x)$ distribution. On the
other hand, the strange antiquark, mainly interacting with
the down quark in $K^0$, yields the $\bar s(x)$ distribution. With
a reasonable choice of parameters (a bag radius of 0.6 fm,
for instance), Signal and Thomas predicted the distributions
 shown in Fig. 2.  Not surprisingly, the two
distributions have quite different shapes, as they
remarked. The $\bar s$ distribution
is relatively soft, like a typical
sea distribution. However, the $s$ distribution
is much harder, resembling the valence
distributions in the nucleon. [A hard distribution
means the large quark density at large $x$, like
valence distributions, and a soft distribution means the
otherwise.] The model predicts that the momentum fraction
of the nucleon carried
by the strange quarks is almost twice as much as that
carried by the strange antiquarks.

If the small locality of the strange sea is confirmed by data,
it might be one of the strongest evidences for
the meson-could model. However, the recent CCFR data
from neutrino deep-inelastic scattering show
that the strange quark and antiquark distributions
are almost identical within the experimental
errors, with $s(x)$ being slightly harder than
$\bar s(x)$: $s(x)/\bar s(x) \sim (1-x)^{-0.46\pm 0.87}$.
The data, analyzed to the next-to-leading
order in the $\overline{ \rm MS}$
scheme, is also shown in Fig. 2\cite{ccfr}.
In a recent paper by Koepf et al.,
the meson-cloud model was used to fit the $\bar s(x)$
distribution by adjusting the cut-off in the
kaon-baryon form factor \cite{koepf}, shown in
the dashed curve in Fig. 2. Although the model
can fit the data points for $x>0.3$, it
underpredicts the data
at smaller $x$. In the same spirit, one might try to fit
the $s(x)$ distribution by adjusting
the bag radius used in Signal and Thomas's calculation
However, according to Ref. \cite{thomas},
not only does one need a substantially large
bag radius, but also the fit works only for
the large $x$ data. In short, it is quite difficult
to find a set of parameters which yield almost
identical $s(x)$ and $\bar s(x)$ distributions
in the meson-cloud model.

Thus the meson-cloud model
for the strange sea has a limited value.
However, it is difficult to understand
from the view of QCD why the
strange distributions have such high degree of
locality. Failed to find a dynamical reason,
we introduce a model which can accommodate varying degree of
locality and use it to fit the strange distributions.
Our intention is to correlate locality in momentum
and coordinate spaces.

We assume the strange pairs
at scale $\mu$ are produced perturbatively
from the gluon distribution $G(x)$ at
the same scale.
To regulate the infrared divergences in the
perturbative calculation, we introduce effective
masses for the strange quarks and antiquarks.
Physically, the effective masses account
for the mean-field interactions between
the sea quarks under consideration and others present
in the nucleon. The different interactions
experienced by the strange quark and antiquarks
are reflected by the difference in the
effective masses. [Note that strictly speaking,
different masses for quarks and antiquarks not only
violate CPT in field theory, but also render
the color current non-conserved. However, at the
level of a model, we ignore these problems.] Intuitively,
when a quark has a large effective mass, the corresponding
distribution is hard.

To simplify the calculation, we work in a
special coordinate system. We assume that
the nucleon is moving in the $z$ direction,
$P = (P^0, 0, 0, P^3)$.
Choose two light-cone
null vectors $p = P^+/\sqrt{2}(1, 0, 0, 1)$ and
$n=1/(\sqrt{2}P^+)(1, 0, 0, -1)$
with $p^2=0$, $n^2=0$, $p\cdot n=1$, where
$P^+ = 1/\sqrt{2}(P^0+P^3)$ is a component of
the nucleon momentum in light-cone coordinates.
The nucleon moment is now
$P=p+\frac {M^2}{2} n$. The quark distribution
in the nucleon can be calculated with
\cite{ji0}
\begin{equation}
  f(x)=\frac 12\int\frac{d\lambda}{2\pi} e^{i\lambda\cdot x}\langle P|\bar\psi
(0)
n\!\!\! / \psi (\lambda n)|P\rangle \ ,
\label{parton}
\end{equation}
where $|P\rangle $ is the nucleon state
normalized as $\langle P|P'\rangle=2P^0(2\pi)^3\delta^3(\vec{P}-\vec{P}')$.

Using the above equation,
we calculate the strange distribution in a gluon,
\begin{equation}
\label{qua}
f_s(x,\mu^2)=\frac{\alpha_s}{4\pi} \int_0^{\mu^2}
dk_{\perp}^2\frac{(k_{\perp}^2+\bar
m_s^2)[x^2+(x-1)^2] + 2x(1-x)m_s\bar m_s+(1-x)^2(
m_s^2-\bar m_s^2)}{[(k_{\perp}^2+\bar m_s^2)+(1-x)(m_s^2-\bar m_s^2)]^2}\ .
\label{x}
\end{equation}
where $m_s$ and $\bar m_s$ are the effective
masses for the strange quarks and antiquarks, respectively,
and $\mu$ is the transverse momentum cut-off defining
the renormalization scale of the distribution.
Since we are going to compare our result with the
experimental data at 1 GeV$^2$, we take
$\mu^2 = 1$ GeV$^2$ accordingly. We choose the
strong-coupling constant, $\alpha_s$, to be
0.5, which roughly corresponds to the
$\overline{\rm MS}$ coupling at the cut-off scale.
The strange antiquark distribution in a gluon
has the similar expression as in Eq. (\ref{x}),
except $x$ and $1-x$ need to be interchanged.

Convoluting $f_s(x,\mu^2)$ with the gluon distribution $G(x,\mu^2)$,
we predict the strange distribution,
\begin{equation}
\label{xs}
xs(x,m_s,\bar m_s)=\int_x^1
{dy\over y} ~xf_s({x\over y},m_s,\bar m_s)~G(y ,\mu^2=1{\rm GeV}^2) \ ,
\end{equation}
Using the CTEQ3 gluon distribution\cite{cteq} and fitting
Eq. (\ref{xs}) to the CCFR data\cite{ccfr},  we get
$m_s=260 \pm 70$ MeV, and $\bar m_s=220 \pm 70 $ MeV,
with large $m_s$ correlated with small $\bar m_s$ and vice versa.
The quality of the fit with $m_s=260$ MeV
and $\bar m_s = 220$ MeV is shown in Fig.~3.
Given the simplicity of the model, the
agreement is remarkably good. The effective masses
$m_s$ and $\bar m_s$ are equal within
the experimental errors. The difference in the
central values, if significant at all, reflects
weakly the points of the meson-cloud model.

Closely related to the momentum-space locality
is the strange radius of the nucleon
introduced by Jaffe in Ref. \cite{jaffe}. Experimentally,
the strange radius can be extracted
from the strange contribution
to the elastic form factor of the nucleon.
Several experiments at CEBAF have been planned
to measure the form factor through
parity-violating electron scattering \cite{musolf}.
Theoretically, the strange radius measures
the net strangeness (strange minus antistrange quark)
distribution in the nucleon (see Eq. (1)).  There
exist several conflicting predictions of the
strange radius in the literature.
The dispersion analysis in Ref. \cite{jaffe}, assuming
the second isoscalar vector-meson pole is dominated by
$\phi$, gave $\langle r^2\rangle_s$ = 0.13 fm$^2$. In
coordinate space, a positive strange radius
means that the strange
quarks dominate over anti-strange quarks at
large radius. On the other hand, the SU(3)
Skyrme model predicted a strange radius in the range of
$-0.11$ to $-0.21$ fm$^2$\cite{park}. The sign here is consistent
with the meson-cloud model. Finally,
the kaon-loop calculations yielded a
strange radius of order $-0.01$ fm$^2$\cite{musolf1}.

The CCFR data seem to imply that the
strange radius is small. To directly
reach this conclusion,
one must find a model to correlate
the strange distributions with the elastic
form factor. [It is not impossible
to find a wavefunction from which the momentum
space locality implies nothing for the coordinate
space locality. However, such a wavefunction seems
unnatural.]  One such model is the particle-plus-core
model suggested by Gunion, Brodsky and
Blankenbecler \cite{gunion} many years ago. For our purpose,
the simplest version without spin
degrees of freedom is qualitatively sufficient.
Using $\psi_s(k_\perp, x)$ to denote the wavefunction for
the strange quarks,
the strange quark distribution is,
\begin{equation}
    s(x) = {1\over 2x(1-x)}\int {d^2k_\perp\over (2\pi)^3}
        |\psi_s(k_\perp, x)|^2 \ .
\end{equation}
On the other hand, the strange form factor is,
\begin{eqnarray}
    F_s(q^2) = \int^1_0 {dx\over 2x(1-x)} \int {d^2k_\perp\over (2\pi)^3}
            && \Big[\psi_{s}(k_\perp, x)\psi_{s}(k_\perp +
       (1-x)q_\perp, x) \nonumber \\
     && - \psi_{\bar s}(k_\perp, x)\psi_{\bar s}
  (k_\perp + (1-x)q_\perp, x) \Big]  \ .
\end{eqnarray}
Thus one can read the wavefunction from an expression for
$s(x)$ and calculate the corresponding elastic form factor.
Almost by definition, the small momentum-space locality
is directly translated to the small coordinate-space locality.
Without an explicit calculation,
we can estimate the strange radius through a dimensional
analysis,
\begin{eqnarray}
     |\langle r^2\rangle_s| &&\sim {(m_s-\bar m_s)\over \mu^3} \nonumber \\
        && \le 0.005 ~{\rm fm }^2 \ .
\end{eqnarray}
Except for an uncertain sign, the bound is smaller than any
of the theoretical prediction mentioned above.

To provide support for the expectation that the momentum space locality may
mean coordinate space locality, we consider the neutron
charge distribution. The valence quark distributions
measured in hard processes directly provide the charge distribution
in momentum space. In Fig. 4, we have shown the positive charge
distribution from the up quark and the negative charge distribution
from the down quarks as fitted in CTEQ3 \cite{cteq}.
The total charge distribution, shown in the solid line,
is considerably smaller than individual charge distributions,
indicating strong locality of charges in momentum
space. From this, one would expect that the charge distribution
in coordinate space is quite local. This is supported
by the small charge radius of the neutron $-$0.12 fm$^2$.
[Notice, however, that the sign correlation
in momentum and coordinate spaces is counter-intuitive.]

One might generalize the notion of ``local similarity"
of the sea to the spin degrees of freedom.
For instance, if quarks and antiquarks tend to have
the same spin wavefunctions, the strange anomalous magnetic
moment might be small. This spin locality is not expected in QCD
even though the spatial degrees of freedom show a high degree
of locality. A heuristic example is the neutron anomalous
magnetic moment, which is large despite the charge
distribution is very much local. However,
QCD dynamics may surprise us again.
The issue can be resolved by measuring the helicity distributions
of the strange quarks ($\Delta s(x)$) and antiquarks $(\Delta \bar s(x)$)
as well as the strange anomalous
magnetic moment $\mu_s$.

Generalizing the concept of locality to all charge-conjugation-odd
observables is tempting but subtle. The charge-conjugation
operation involves not only the interchange of $s$ and $\bar s$
quarks, but also the change of the sign for the gluon fields.
The nucleon is definitely very different if all
gluon fields are charge-conjugated.
However, for observables and/or models
without the explicit gluon fields, one can discuss
the generalized locality through the observation
that all charge-conjugation-odd observables
vanish in a state with definite charge-conjugation properties.
In fact, if one defines a charge-conjugation operation
as merely the interchange of $s$ and $\bar s$ quarks
and if the sea has definite charge-conjugation symmetry,
the matrix elements of C-odd operators with strange
fields only, such as $\langle PS|\bar
s \sigma^{\mu\nu} s|PS\rangle$ and $\langle
P'|\bar s \gamma^\mu s|P\rangle$, vanish.
In a phenomenological model proposed by Jaffe and
Lipkin to account for the spin properties of the nucleon
\cite{lipkin}, the sea quarks form $0^{++}$ and
$1^{++}$ pairs. Here the sea is charge-conjugation even.

The CCFR data provide a strong evidence
that quarks and antiquarks in the strange sea encounter
similar interactions. It is important to
check independently that the CCFR method
of extracting the strange distributions actually
works. In deep-inelastic scattering, there is
another way to measure the strange distribution
through combining the structure function $F_2(x)$
measured from muon and neutrino scatterings
\cite{nmc}.
In a recent paper by Barone
{\it et al.} \cite{barone}, they
analyzed both methods by taking into account the quark mass
and current non-conservation effects.
They concluded that to the next-to-leading order,
both extractions of the strange distributions
are consistent.
Hadron facilities provide new opportunities to access
to the strange distributions in the nucleon. Intensive kaon beams,
if available, can probe the strange sea directly through
Drell-Yan processes\cite{na3}. The polarized RHIC has the
capability of selecting hard, weak-interaction
subprocess through single-spin asymmetry, and thus
strange quarks, with their unique weak coupling, can be
identified through their transition to charm quarks\cite{soffer1}.
Clearly, the polarized and unpolarized strange distributions
deserve to be studied more extensively
in the future experiments.

To summarize, we introduced the concept of ``locality'' for
the strange sea in the nucleon, which measures the local
similarity of strange quark and antiquarks in momentum
and coordinate spaces.
We pointed out that the CCFR data on the strange distributions
indicate that the strange sea is highly local,
a picture unexpected in QCD and at variance with the
meson-cloud model. We introduced a
simple perturbative model for the strange sea,
which fits the CCFR data very well. In our model,
locality of the sea is reflected by the difference
of the effective masses for the strange quarks and antiquarks.
By correlating the strange
distribution with the elastic form factor, we argue
that the strange radius of the nucleon is
smaller than 0.005 fm$^2$, a prediction to be
verified soon by CEBAF experiments.
Finally, we discussed the extension of the locality concept
to other observables and commented on the experimental
status of the strange distributions.

\acknowledgments

We wish to thank D. Beck, M. Burkardt, R. Jaffe,
J. Negele, and J. Soffer for many useful discussions.

\begin{figure}
\bigskip

\caption{Feynman diagrams representing interactions of
the sea quark and antiquark in a pair with
valence quarks in the nucleon.}
\label{fig1}

\bigskip
\caption{The meson-cloud model predictions for
the strange and anti-strange distributions in the nucleon and
the CCFR data. The solid lines are from Ref. [5] and the dashed
line from Ref. [8].}
\label{fig2}

\bigskip
\caption{Comparison of our model predictions ($m_s = 260$ MeV
and $\bar m_s = 220$ MeV) with the CCFR data.}
\label{fig3}

\bigskip
\caption{The neutron charge distributions
as functions of the Feynman $x$. The positive charge $\rho_+$
is from the valence $u$ quark and the negative charge $\rho_-$
from the valence $d$ quarks.}
\label{fig4}

\end{figure}

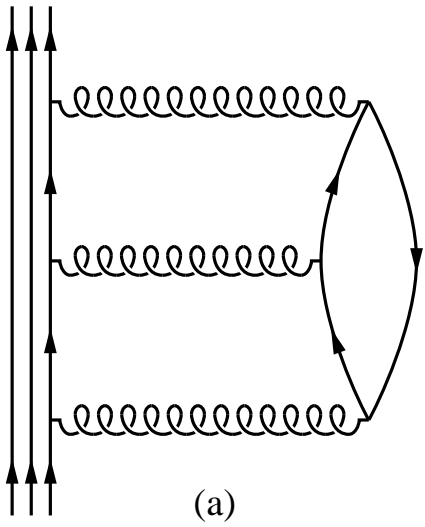
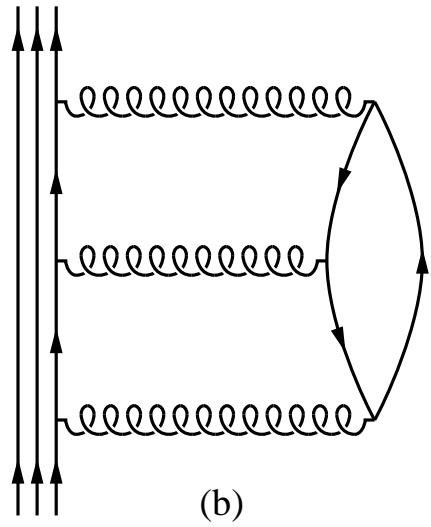
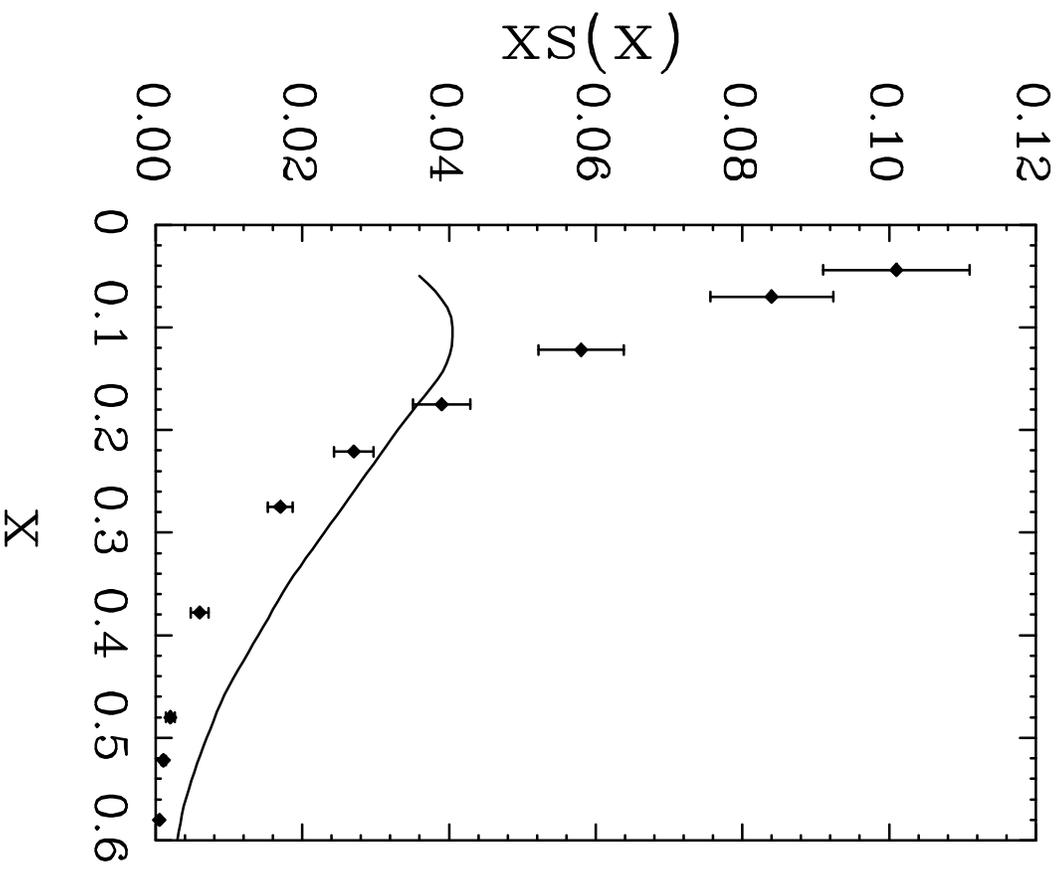
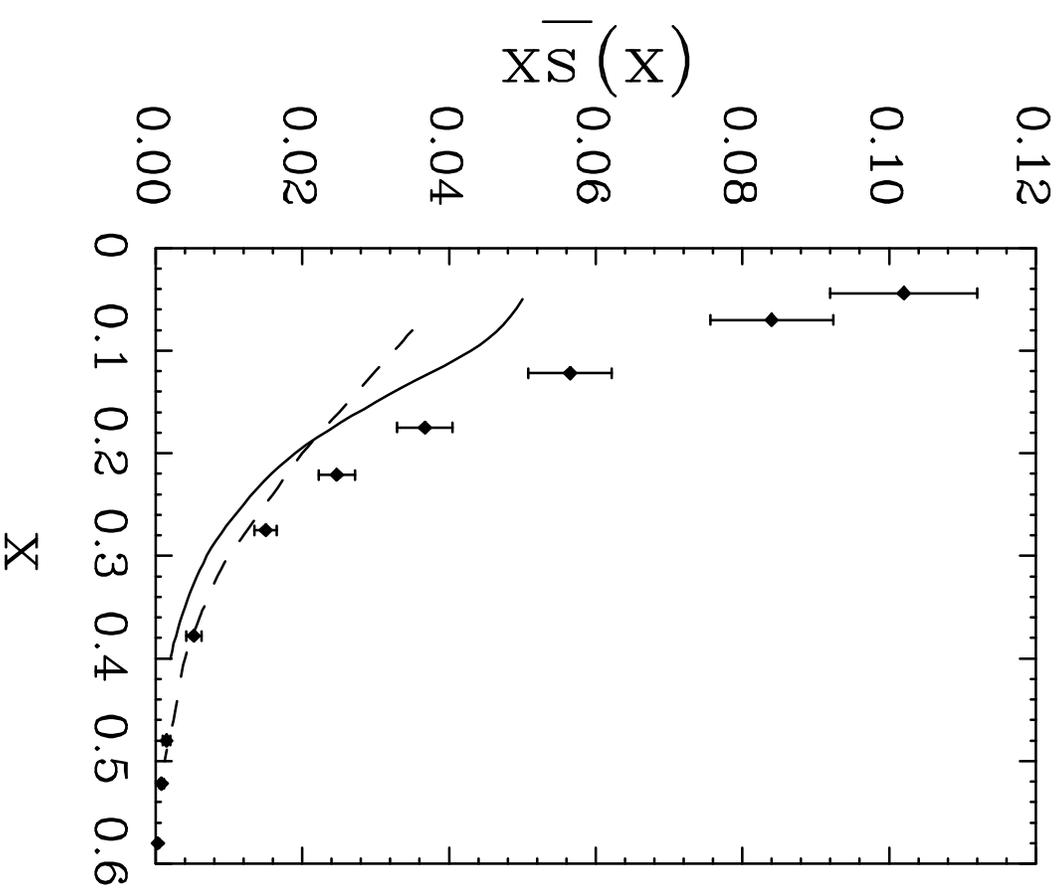
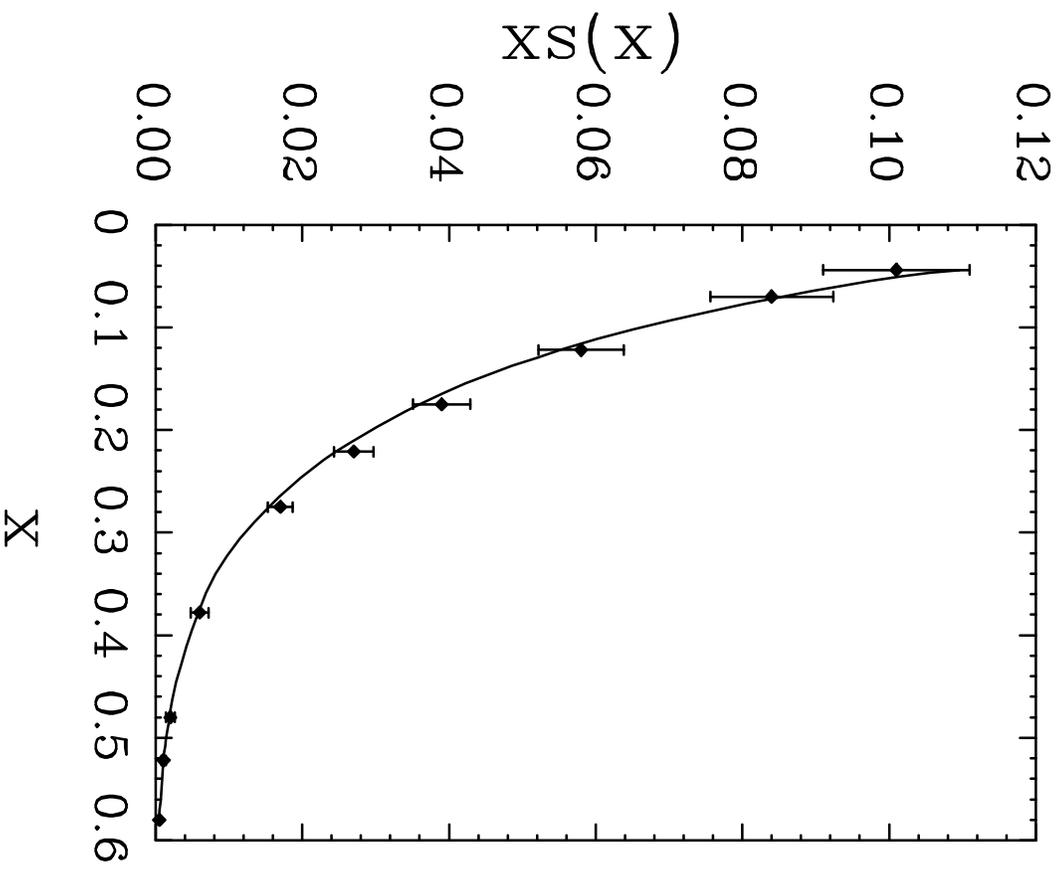
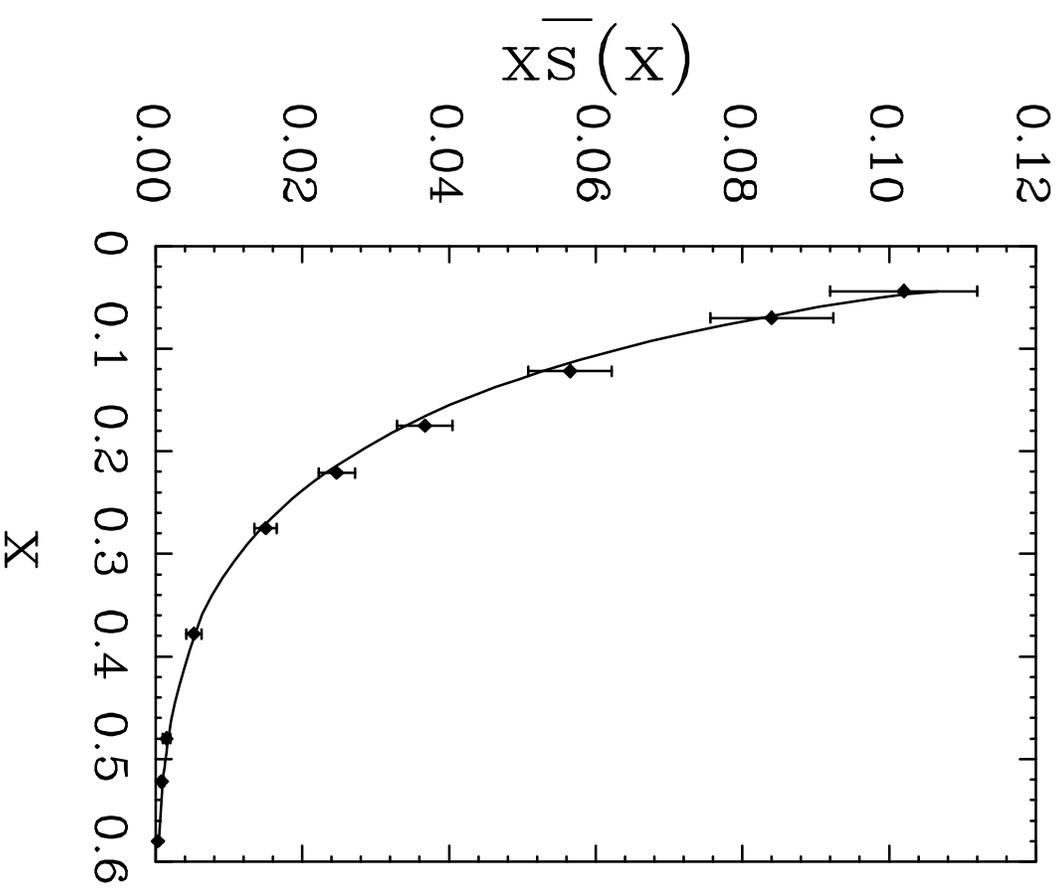
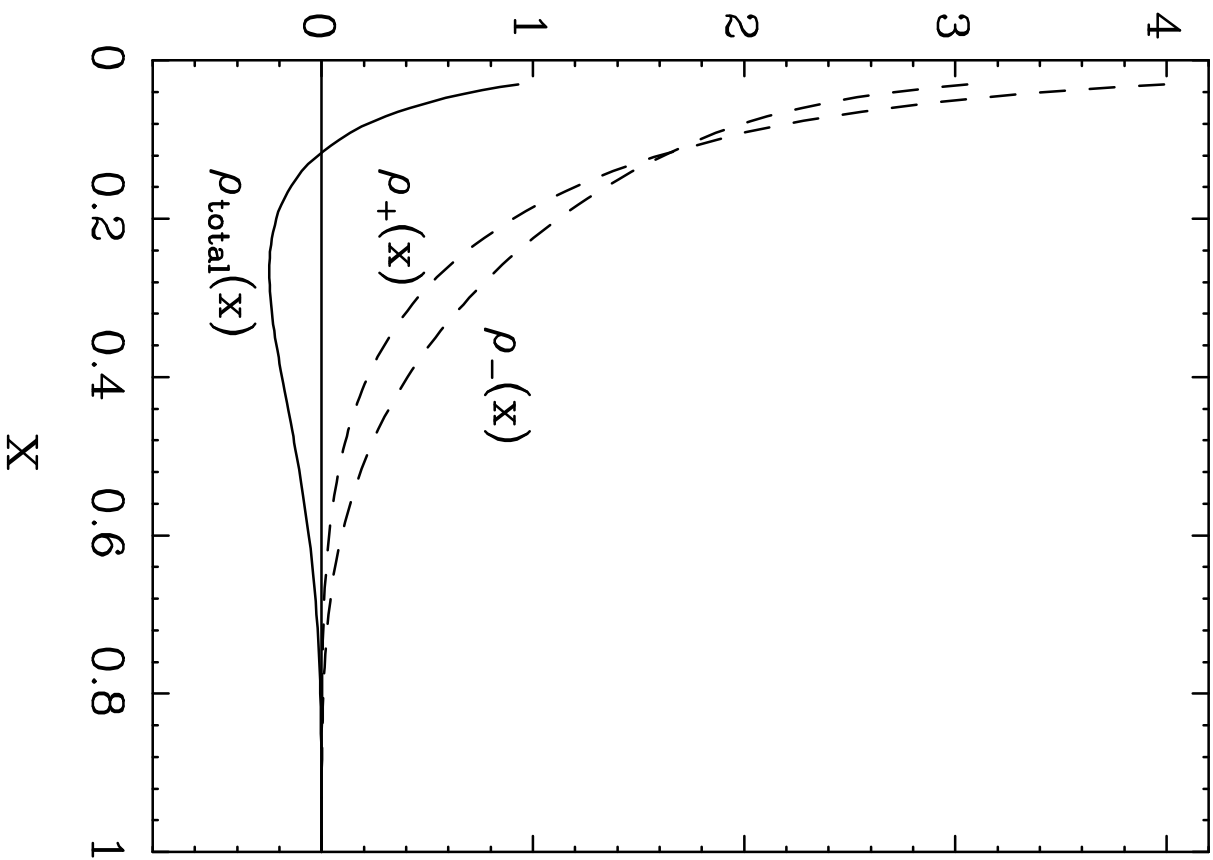
\end{document}